\documentclass[]{emulateapj-rtx4}

\usepackage{natbib,color} 
\bibpunct{(}{)}{;}{a}{}{,}

\DeclareRobustCommand{\ion}[2]{%
\relax\ifmmode
\ifx\testbx\f@series
{\mathbf{#1\,\mathsc{#2}}}\else
{\mathrm{#1\,\mathsc{#2}}}\fi
\else\textup{#1\,{\mdseries\textsc{#2}}}%
\fi}

\shorttitle{Fast NLTE Inversion of Solar \ion{Ca}{ii} Spectra}
\shortauthors{Beck, C.; Sanjay, G.; Kiessner, C.}

\begin{document}
\title{Fast Inversion of Solar \ion{Ca}{ii} Spectra in Non-Local Thermodynamic Equilibrium} 


\author{C. Beck}
\affil{National Solar Observatory (NSO),  Boulder}
\author{S. Gosain}
\affil{National Solar Observatory (NSO), BoulderA}
\author{C. Kiessner}
\affil{University of Hawai'i at Hilo}


%\altaffiltext{1}{Departamento de Astrof\'isica, Universidad de La Laguna, E-38205 La Laguna, Tenerife, Spain}


\begin{abstract}
Present day solar imaging spectrometers typically yield a few hundred million spectra in one hour of observing time. This number will increase by an order of magnitude for future instruments with larger 4k\,$\times$\,4k sensors as planned to be used for the upcoming Daniel K.~Inouye Solar Telescope. A fast quantitative analysis of such huge data volumes can be done by comparing the observations to an archive of pre-calculated synthetic spectra to infer the thermodynamic properties of the atmosphere. To analyze intensity spectra of the \ion{Ca}{ii} IR line at 854\,nm in the solar atmosphere, we generated an archive with 2,000,000 spectra under the assumption of non-local thermodynamic equilibrium (NLTE) with the NICOLE code \citep{socas-navarro+etal2015}. We tested its performance by inverting 60 spectral scans of \ion{Ca}{ii} IR at 854\,nm in the magnetically quiet Sun with 700,000 profiles each. Based on the inversion results obtained using the full archive, we construct a smaller archive by keeping only the about 70,000 archive profiles that were actually used. We can reproduce the observed intensity spectra to within a few percent using either the full or the small archive. For spectra with 30 wavelength points, this NLTE inversion approach takes 0.02 (0.35)\,s per profile  to obtain a temperature stratification when using the small (full) archive, i.e., it can invert a single spectral scan in about 4 (68) hrs. The code is able to simultaneously deal with an arbitrary number of spectral lines. This makes it a promising tool for deriving thermodynamic properties of the solar atmosphere from current or future solar high-resolution observations of photospheric and chromospheric lines. 
\end{abstract}

\keywords{line: profiles -- methods: data analysis -- Sun: chromosphere \\{\it Online-only material:\rm} color figures}
\section{Introduction}
In contrast to night-time observations of stars that lack spatial resolution, observations of the Sun as our closest star provide an unprecedented amount of spatially and temporally resolved information. Acquiring solar data with high spectral resolution adds resolved information in a third dimension, i.e., in height in the solar atmosphere. With spectropolarimetric observations one can derive the properties of solar magnetic fields in addition to thermodynamic quantities that are accessible from plain spectroscopy. 

The observational requirements for a full inference of physical parameters in the solar photosphere and chromosphere are to resolve the relevant spatial structures on the solar surface (spatial resolution of 0\farcs1 or better), the thermal broadening of spectral lines at temperatures of 6000--20,000\,K (spectral resolution of 2-10\,pm) and the characteristic time scales of the evolution of the atmosphere (cadence of 1--100\,s). Modern solar instrumentation is able to fulfill some or all of these requirements at the same time \citep[e.g.,][]{beck+etal2005b,socasnavarro+etal2006,cavallini2006,gosain+etal2006,collados+etal2007,scharmer+etal2008,tsuneta+etal2008,jaeggli+etal2010,jess+etal2010,puschmann+etal2012b,depontieu+etal2014,bjorgen+etal2018}. With the rapid technical progress of sensors commonly used in solar instrumentation both in detector size and frame rates, a massive problem has, however, surfaced in recent years: the sheer data volume. At a frame rate of about 10\,Hz, imaging spectrometers can record a single spectral scan of a solar spectral line in about 5--10\,s \citep{iglesias+etal2016}. This typically yields data sets with a few hundred million spectra per hour of observations for a 1k\,$\times$\,1k sensor, which requires fast analysis methods for their evaluation in order to keep up with the observations. 

%As solar observations are full of relevant information, one wants to extract as much as possible of it from the observed spectra in a quantitative analysis. 
%One wants to . 

The common approach for extracting as much information as possible from observed solar spectra in a quantitative analysis is to subject them to a so-called ``inversion''. As there is no direct way to convert an observed intensity or polarization signal into a thermodynamic or magnetic property at some height in the solar atmosphere, the inverse problem of finding the solar atmosphere model that best reproduces the observations is solved instead \citep[e.g.,][]{iniesta+cobo2016}. A variety of inversion codes exists for the solar case. Their main difference is the complexity in dealing with the radiative transfer equation (RTE) ranging from analytical solutions in the Milne-Eddington approximation \citep{borrero+etal2011}, solving the RTE assuming local thermodynamic equilibrium \citep[LTE; e.g.,][]{cobo+toroiniesta1992} to finally using non-local thermodynamic equilibrium  \citep[NLTE; e.g.,][]{socas-navarro+etal2015}. The main driver for increasing the complexity and realism of the treatment of the RTE is that in the solar chromosphere the assumption of LTE starts to break down because the number of particle collisions is insufficient to distribute the energy accordingly over all available degrees of freedom. One major consequence of the departure from LTE is a de-coupling between the local gas temperature and the radiation field that are no longer coupled directly through the Planck function \citep[e.g.,][]{rutten2003}. This naturally has a strong impact on the analysis of observations that correspond to the radiation field integrated over the atmosphere. 

A fast inversion method for chromospheric spectra of \ion{Ca}{ii} lines in LTE was introduced in the past \citep{beck+etal2013,beck+etal2015}. It was successfully applied to observational data with or without an additional approximate NLTE correction \citep{beck+etal2013a,beck+etal2014,rezaei+beck2015,choudhary+beck2018,grant+etal2018}. In the current study, we analyze the performance of a fast NLTE inversion based on an archive of pre-calculated spectra synthesized with the Non-LTE Inversion COde using the Lorien Engine \citep[NICOLE;][]{socas-navarro+etal2015}. Section \ref{sec_obs} describes the observations used for testing the archive. Section \ref{nlte_inv} details the NLTE inversion process. The results are given in Section \ref{results} and summarized in Section \ref{summsumm}, respectively. We discuss our findings in Section \ref{disc}. Section \ref{concl} provides our conclusions.

\section{Observations}\label{sec_obs}
\begin{figure}
\centerline{\resizebox{8.8cm}{!}{\includegraphics{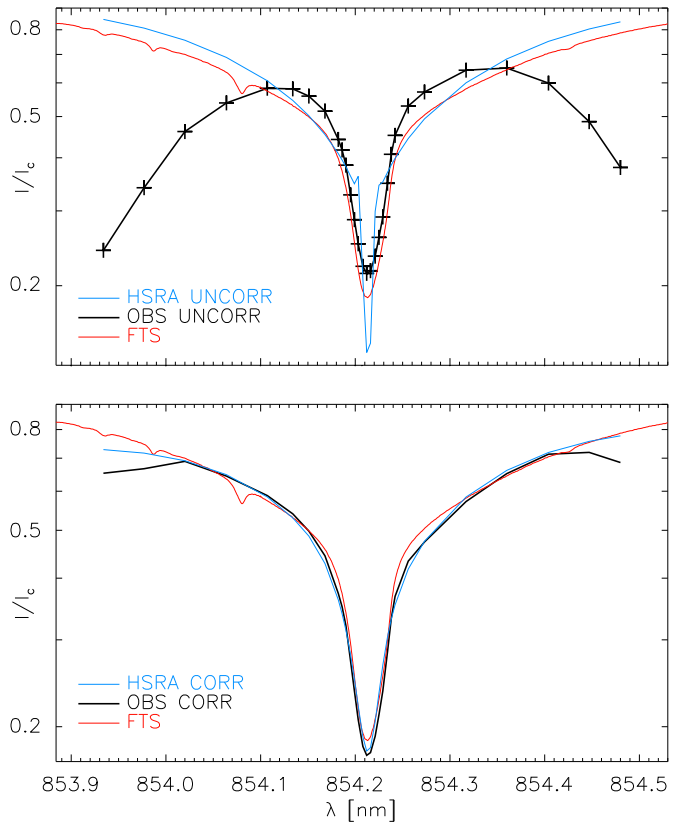}}}
\caption{Matching of spectra. Top panel: FTS reference spectrum (red line), uncorrected average observed spectrum (black line) and original HSRA NLTE spectrum (blue line). Bottom panel: FTS reference spectrum (red line), corrected observed spectrum (black line) and matched HSRA spectrum (blue line).} \label{fig1}
\end{figure}

The data used here consist of a time-series of 396 spectral scans of the \ion{Ca}{ii} IR line at 854.2\,nm acquired with the Interferometric BiDimensional Spectropolarimeter \citep[IBIS;][]{cavallini2006} at the Dunn Solar Telescope on September 18, 2015, from UT 14:10 until 15:25. The \ion{Ca}{ii} IR line was scanned on 30 non-equidistant wavelength positions from 853.95\,nm to 854.45\,nm (top panel of Figure \ref{fig1}). The cadence was about 10\,s because of sequentially scanning the H$\alpha$ line in between. The spatial sampling was about 0\farcs095\,pixel$^{-1}$ with a total circular field of view (FOV) of 95$^{\prime\prime}$ diameter. The FOV was located at disc center in a quiet Sun (QS) region devoid of strong magnetic activity. 

The data were reduced with the standard IBIS data reduction pipeline\footnote{\url{https://www.nso.edu/telescopes/dunn-solar-telescope/dst-pipelines/}}. Before the execution of the inversion, three additional steps were applied to the spectra. The effect of the transmission curve of the prefilter (top panel of Figure \ref{fig1}) was removed; the average profile over a 600$\times$600 pixel region around the center of the FOV was normalized to the intensity of the reference spectrum from the Fourier Transform Spectrometer atlas \citep[FTS;][]{kurucz+etal1984} in the blue line wing; and finally, a residual trend in the wavelength direction was taken out by division with a straight line to roughly equalize the intensities in the blue and red line wing. The application of these three corrections to all observed spectra yielded the corrected average observed profile (black line in bottom panel of Figure \ref{fig1}) that maintains its intensity  normalization to the FTS atlas spectrum.

\section{Non-local Thermodynamic Equilibrium Inversion}\label{nlte_inv}
The application of the spectral archive assuming NLTE follows to a large extent the description given in \citet{beck+etal2013} and \citet{beck+etal2015} for the LTE version with minor modifications.
\subsection{Generation of Spectral Archive}
The NLTE spectral archive was generated by synthesizing spectra from a series of model atmospheres created by introducing perturbations to the temperature stratification of the Harvard-Smithsonian Reference Atmosphere \citep[HSRA;][]{gingerich+etal1971}. The main advantage of using NICOLE over the Stokes Inversion Based on Response functions \citep[SIR;][]{cobo+toroiniesta1992} LTE code for generating the spectral archive is that with NICOLE the basic chromospheric temperature rise above log\,$\tau \approx -4$ does not need to be removed or mitigated before the spectral synthesis because it does not lead to excessive emission as in the LTE case. 

For the creation of the NLTE archive, we added the following temperature perturbations to the HSRA model in a similar way as for the LTE case:
\begin{itemize}
\item[1.] A global temperature offset $\Delta T$ at all optical depths, from $-1000$ to $+200$\,K in steps of $20-40$\,K. 
\item[2.] Localized Gaussian perturbations with the free parameters amplitude $A$ ($10-100$\,K in 10\,K steps), width $\sigma$ ($0.2-3.1$ in units of log\,$\tau$ in steps of 0.1) and location in log\,$\tau$ ($-8$ to $+1.4$ in steps of 0.1).
\end{itemize}
The amplitude $A$ of the Gaussian perturbation was defined as 10--100\,K at log\,$\tau = 0$, but then was scaled up with the location of the Gaussian in optical depth to be larger in the upper atmosphere \citep[see Figure 2 of][]{beck+etal2013}. The scaling factor was $\approx$\,30 at log\,$\tau = -8$, i.e., an initial value of $A=100$\,K at log\,$\tau = 0$ converts to $A=3000$\,K at log\,$\tau = -8$.

Unlike the LTE case, we ran the spectral synthesis with both the addition and subtraction of the Gaussian perturbations to the HSRA. In the LTE case, the modified HSRA atmosphere, especially for negative temperature offsets, already corresponds to sort of a lowermost temperature to ever be expected. For the NLTE case, the chromospheric temperature rise in the HSRA cannot be removed in a similar way by the global offsets $\Delta T$. Stability reasons in solving the NLTE equations require a minimal temperature in the stratification of about 2000\,K, which prevents using global offsets $\Delta T$ below about $-2000$\,K. To reduce the chromospheric temperature rise by larger amounts could thus only be done using negative Gaussian temperature perturbations. We iterated the generation and application of the archive to a single randomly picked spectral scan a few times to fine-tune the exact settings of the archive generation until the current result was achieved. 

The resulting about 2,000,000 temperature stratifications were then put into hydrostatic equilibrium in NICOLE and the corresponding spectra of the \ion{Ca}{ii} IR line at 854.2\,nm were synthesized. The spectral sampling was 2\,pm with 500 spectral points over a spectral range from 853.7 to 854.7\,nm. The synthesis takes about 17\,ms per profile on a 32-core desktop machine and thus about 9\,hrs for the full archive. We note that any spectral line that NICOLE can deal with can be synthesized from the archive of temperature stratifications for applications to other lines or combinations of lines, which we plan to explore in future work. 

\subsection{Preparation of Archive for Inversion}
The spectral synthesis of the archive spectra is done with a fine spectral sampling so that it can be degraded to match different observations with varying spectral resolution. No macro- or microturbulence is used in the synthesis leading to very narrow absorption profiles (top panel of Figure \ref{fig1}). Both factors imply that the initial spectral archive has to be matched to the actual properties of the observational data, i.e., spectral sampling and spectral resolution of the instrument used, and any additional broadening of solar origin has to be still accounted for.  

To match the archive spectra to the generic properties of the observations the average observed quiet Sun profile as described above and the synthetic spectrum corresponding to the unperturbed HSRA model are used. The best implementation for this process is unfortunately to some extent variable and not necessarily fully physically motivated, depending on the characteristics of the observational data. The sort of generic list of steps employs a spectral convolution of the initial HRSA archive spectrum with a Gaussian of some width $\sigma$ to match the observed line width. Depending on the difference of the line depth in the convolved HSRA archive spectrum and the average observed profile, an additional correction for the line depth by a constant straylight offset $\alpha$ is usually needed. While the presence of parasitic straylight \citep{cabrerasolana+etal2007,beck+etal2011a} in the real world can only reduce the line depth, matching the convolved HSRA spectrum to the observations may require one to increase the line depth instead, i.e., one has assume a negative straylight contribution if the required broadening $\sigma$ is large and has reduced the line depth too much. 

For the IBIS data sets used, it was not possible to get a good match of the degraded HSRA profile to the average observed profile in the line wing and the line core at the same time using only $\sigma$ and $\alpha$. We thus used these two parameters to primarily match the line core and added an independent correction by two straight lines to match the line wings. With that approach, the degraded and corrected HSRA archive profile matches the observed average QS profile very closely apart from the first and last two wavelength points (bottom panel of Figure \ref{fig1}). The somewhat arbitrary, brute-force degradation that mimics the degrading instrumental effects led us to call the corresponding values of $\sigma$ and $\alpha$ the magic numbers, but it has no real impact on the inversion itself. The only requirement for a successful matching is that the degraded version of the synthetic HSRA profile reproduces the average observed profile, regardless how this is achieved. The match implies that the average QS profile would yield the HSRA temperature stratification when inverted, and that averaging the inversion results over a large QS area to first order also has to yield the HSRA stratification again. 

We note that this is also a limitation of the archive inversion approach because all determined temperatures are derived in some sense as deviations from the HSRA model. It is thus not possible to independently determine, e.g., the average temperature of the quiet Sun at log\,$\tau = 0$ because the corresponding value of the HSRA should result by default. To improve the approach to retrieve absolute temperatures without any specific reference requires to get a realistic line width and line depth in the archive spectra already automatically in the spectral synthesis, for which, however, \emph{a priori} knowledge of solar macro- and especially height-dependent microturbulence would be needed. The match to the spectral resolution of the data should then also be constrained using measurements or theoretical values for the spectral resolution and the amount of parasitic straylight in the instrument. 

\begin{figure}
\centerline{\resizebox{8cm}{!}{\includegraphics{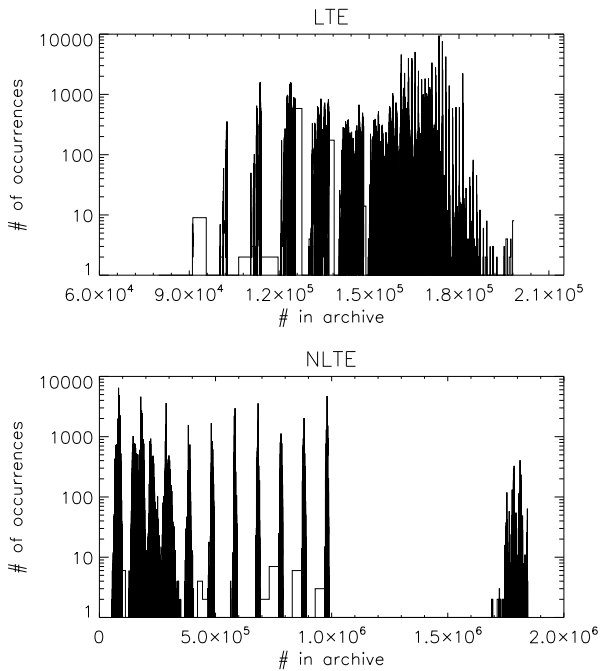}}}
\caption{Archive usage for the LTE (top row) and NLTE archive (bottom row) in a single spectral scan. The x-axis gives the index inside the archive, the y-axis how often that profile was used. The total number of profiles in one spectral scan is about 700,000.} \label{fig4}
\end{figure}

\begin{figure*}
\centerline{\resizebox{17cm}{!}{\includegraphics{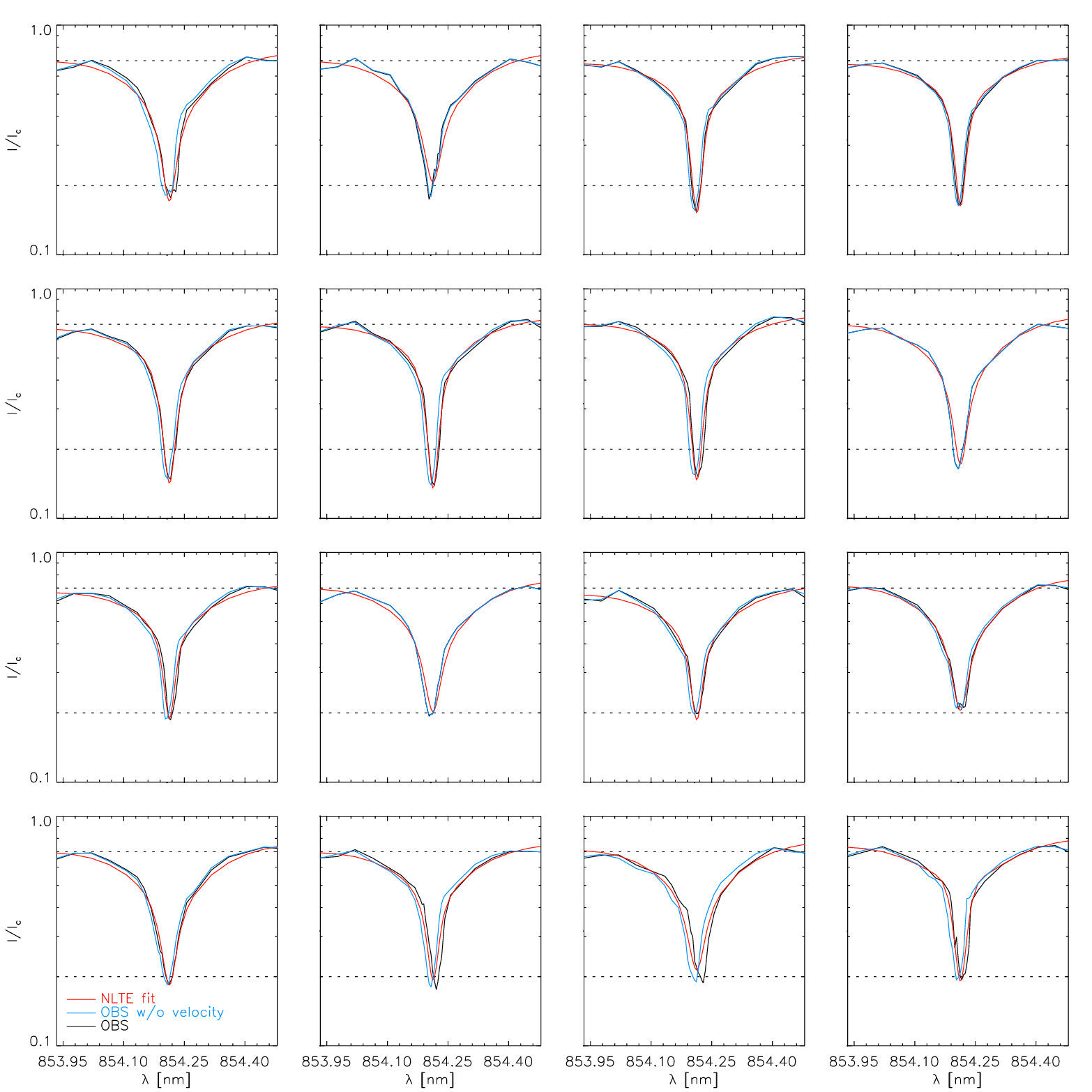}}}$ $\\$ $\\
\caption{Observed and NLTE best-fit spectra from 16 randomly selected locations. Black lines: observed spectra including actual Doppler shifts. Blue lines: observed spectra without Doppler shifts. Red lines: NLTE best-fit spectrum, derived ignoring the Doppler shifts.} \label{fig9}
\end{figure*}
When the needed spectral degradation was identified through testing it on the synthetic HSRA spectrum, the same degradation steps were applied to all archive profiles.
% The degraded archive needs, however, not be stored at the full spectral resolution, because the last step of the matching procedure is to meet the actual spectral sampling of the observations.
For our IBIS data, the spectral re-sampling of the archive from 500 to 30 wavelength points reduced the size of the spectral archive by more than one order of magnitude. The same degraded and re-sampled archive spectra could then be applied to the full time-series because the spectral characteristics of the observations did not change significantly during the data acquisition. Application to observational data with fairly different spectral properties or just a different spectral sampling requires to generate a new degraded archive version, which is a rather fast process of a few minutes, while the initial archive at full resolution is only synthesized once. 

\subsection{Inversion Procedure}
Prior to the actual inversion, both the observed spectra and the degraded archive are multiplied with the weights in the spectral dimension. We set the weights to the inverse of the average observed profile. This choice increases the contribution of the line core to the $\chi^2$ value, which is defined in the standard way as the squared difference between the observed and the archive profiles. The inversion procedure then reduces to comparing each observed spectrum to the whole archive, calculating the $\chi^2$ value, and selecting the archive profile that yields the minimal $\chi^2$. The index of the best-fit spectrum in the archive is used to retrieve the corresponding temperature stratification and all other atmospheric thermodynamic parameters. Instead of looking up the best-fit temperature, one can also only store the best-fit index value to reduce memory usage during the inversion process. There is no specific search algorithm that is used inside the archive, as it is to some extent created in a random order. The best-fit solution automatically corresponds to the global minimum of $\chi^2$ throughout all of the archive. The search for the best match is the primary constraint for the timing that depends most strongly on the size of the spectral archive and the number of wavelength points.
\begin{figure*}
\centerline{\resizebox{7cm}{!}{\includegraphics{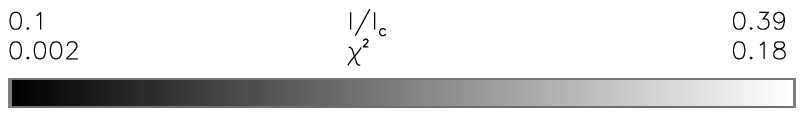}}}$ $\\
\centerline{\resizebox{14cm}{!}{\includegraphics{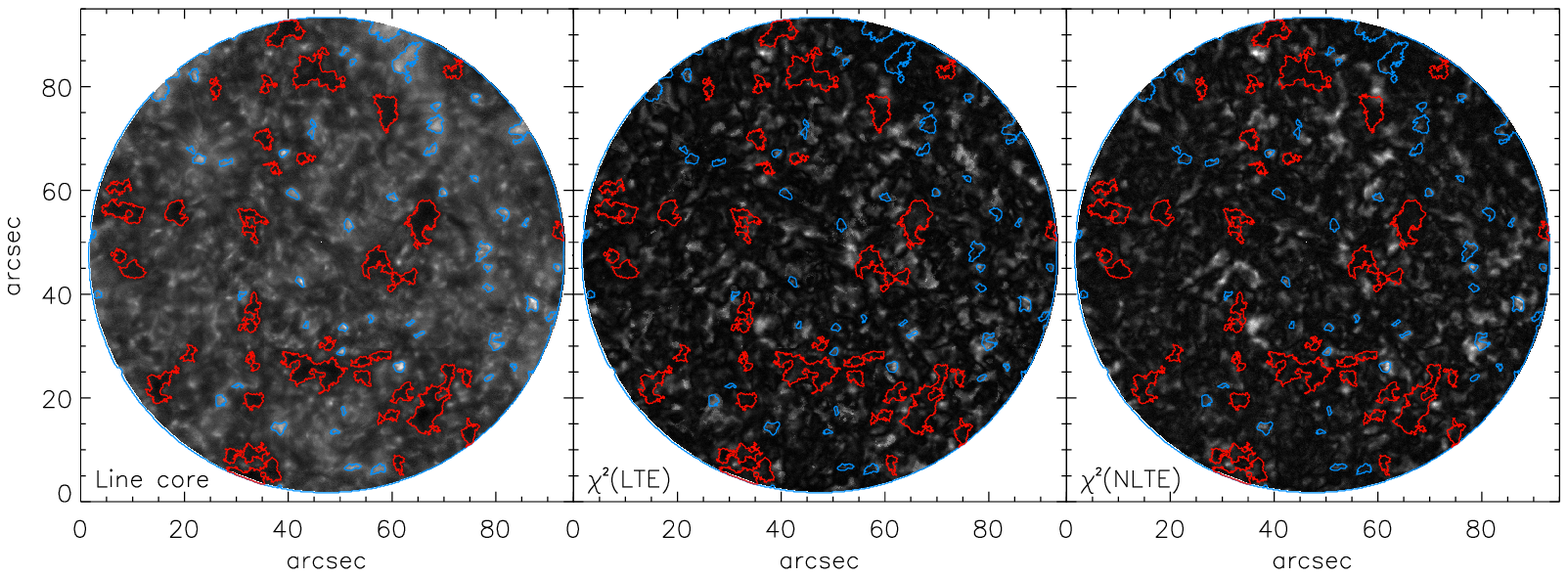}}}$ $\\$ $\\
\caption{Line-core intensity image (left panel) and $\chi^2$ values across the FOV. The 2nd (3rd) panel shows the $\chi^2$ values from the LTE (NLTE) inversion. } \label{fig2}
\end{figure*}

\begin{figure}
\centerline{\resizebox{8.8cm}{!}{\includegraphics{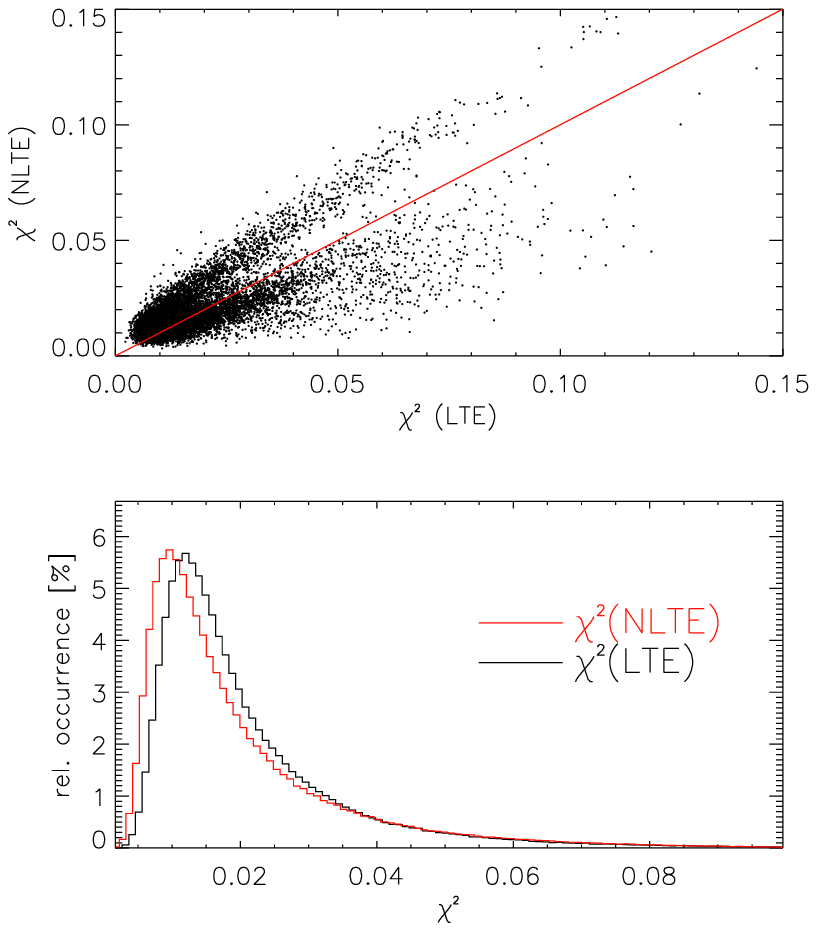}}}
\caption{Comparison of $\chi^2$ in the LTE and NLTE inversion. Top panel: scatterplot of $\chi^2$ in the NLTE inversion vs.~the LTE results. Bottom panel: histograms of the $\chi^2$ values for the LTE (NLTE) inversion in black (red) lines.} \label{fig3}
\end{figure}
\section{Results}\label{results}
\subsection{Computational Requirements and Performance}
The majority of the inversions were done on a quad-core desktop computer with 32\,GB of RAM and 3.5 MHz CPU speed running IDL under Linux. The IBIS data used have a size of 125\,MB per spectral scan, the spectral archive at full resolution and the corresponding atmospheric parameters have about 7.4\,GB, while the degraded archive spectra are about 900\,MB. A single inversion job thus requires about 10\,GB of RAM when the atmospheric model is kept in memory to instantly retrieve the best-fit temperature. The IDL implementation of the inversion routine uses matrix operations where possible to make full use of the capabilities of IDL that parallelizes automatically when its external matrix operation routines are called. As long as no other jobs compete for the CPU power, a speed of about 0.3\,s per profile resulted for the NLTE archive of 2,000,000 spectra and 30 wavelength points. Tests on a 32-core machine gave the same maximal speed of 0.3\,s per profile for the NLTE archive with the only advantage that multiple jobs can be run at the same time at this speed before the CPU power is fully used up. The much smaller LTE archive of 240,000 spectra scaled almost linear with the archive size and gave 0.03\,s per profile. The LTE inversion of all 396 spectral scans was already available prior to the current study and is used for cross-comparisons in the following. Running the LTE version over full-disk \ion{Ca}{ii} IR spectra with 100 wavelength points from the Synoptic Optical Long-term Investigations of the Sun \citep[SOLIS;][]{keller+etal2003} gave 0.1\,s per profile \citep{beck+etal2018}. All steps after the spectral synthesis are conceptually simple and straightforward. They could be implemented in any other programming language as well. The only step that requires optimization for speed is the search for the minimal $\chi^2$ across the archive.

The archives are created without \emph{a priori} knowledge of the observations. They contain an as large as possible variety of different spectra and thermal stratifications, but the full range is not necessarily needed to reproduce the observations. This is especially true for observations in the quiet Sun, where to first order all granules are "similar" to some degree with a hot center surrounded by cool downflows. It turned out that the archive use in the inversion is rather sparse \citep[Figure \ref{fig4}, see also][]{beck+etal2014}. For the same IBIS spectral scan, the LTE (NLTE) inversion only picked $\approx$\,10000 (24,000) different profiles. For the NLTE case, the majority of the profiles corresponded to temperature stratifications with positive Gaussian temperature perturbations in the first half of the NLTE archive (lower panel of Figure \ref{fig4}). 

To confirm the sparse archive usage, we ran the NLTE inversion over the first 60 IBIS scans of the time-series with an average of 20,000$\pm$4,000 profiles used. Combining the inversion results of those scans gave a total of about 70,000 unique profiles that were used in one or the other spectral scan. Thus, only a small fraction of the archive profiles is used, which implies that further optimization is easily possible (see Section \ref{sec_small} below). 

\begin{figure*}
\centerline{\resizebox{14cm}{!}{\includegraphics{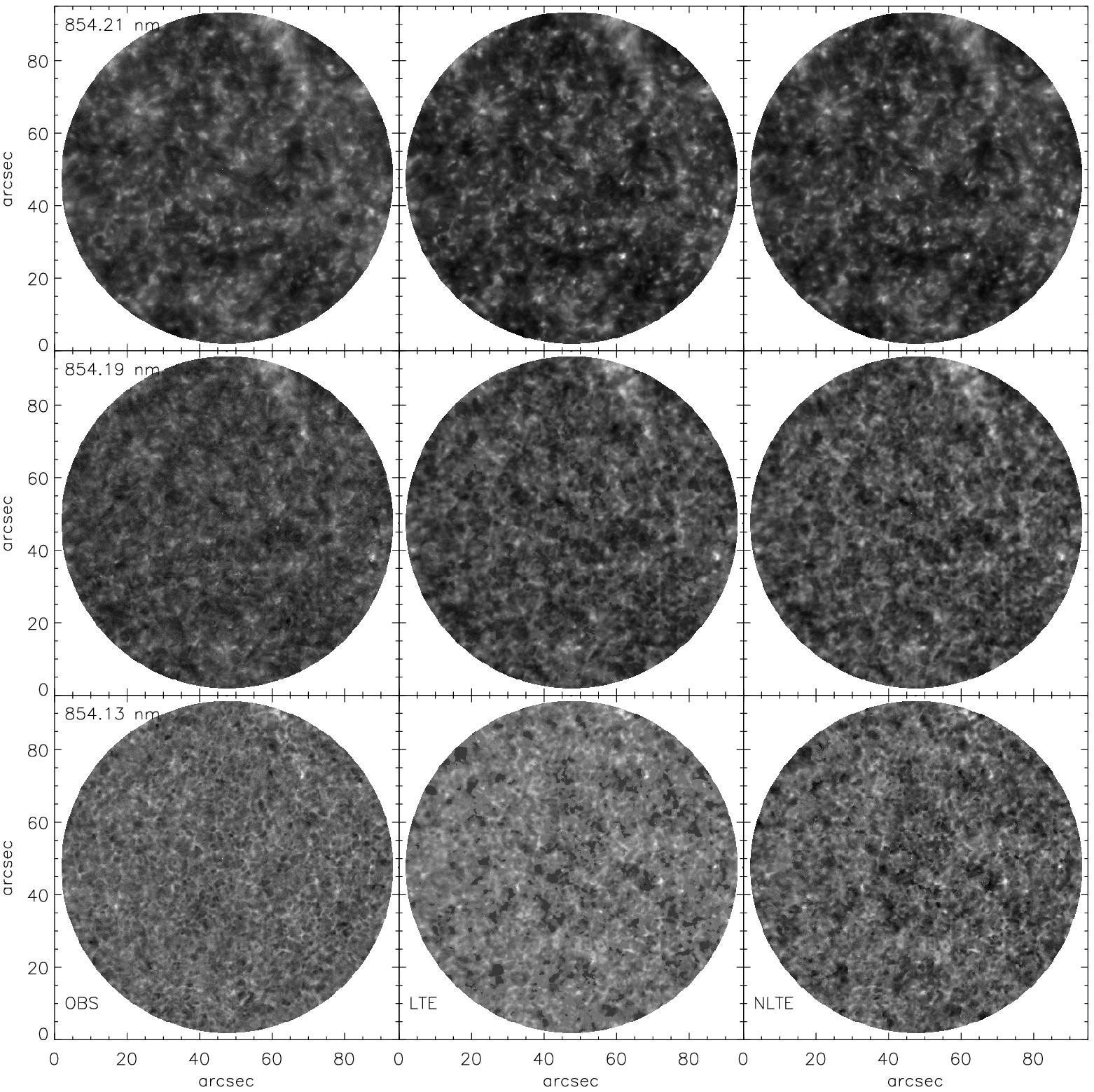}}}$ $\\$ $\\
\caption{Comparison of 2D images from the spectral scan No.~252 at three wavelengths. Left column: observed spectra with Doppler shifts removed. Middle and right column: best-fit spectra from the LTE and NLTE inversion, respectively. The wavelength is indicated in the upper left corner of the panels in the first column.} \label{fig6}
\end{figure*}

\subsection{Impact of Doppler Velocities}
The archive spectra are matched to the average observed QS profile, which also sets the wavelength scale and the zero position of the \ion{Ca}{ii} IR line core to be used. Additional spectral displacements by Doppler shifts due to motions relative to the solar surface are not explicitly taken into account. It is not possible to include Doppler shifts into the archive since that would increase its size multiple times. Fortunately, line-of-sight (LOS) Doppler velocities have only a small impact on the fit for two main reasons. The typical LOS velocities of a few km\,s$^{-1}$, with their maximal value limited by the sound speed of about 6\,km\,s$^{-1}$, lead to spectral displacements of a few pm, which correspond to about two wavelength points at the finest spectral sampling around the line core used in the observations. In addition, the observed and synthetic spectra are to first order symmetric. This forces the best-fit solution to also to first order ignore the Doppler displacement because the mismatches in the blue and red line wing go into opposite directions. If the observed profile is displaced from the location in the archive, the intensity in one wing will be too high, while it will be too low in the opposite wing. The best match then to first order is the same as in the absence of Doppler shifts with minimal deviations in both line wings. 
\begin{figure*}
\centerline{\resizebox{16cm}{!}{\includegraphics{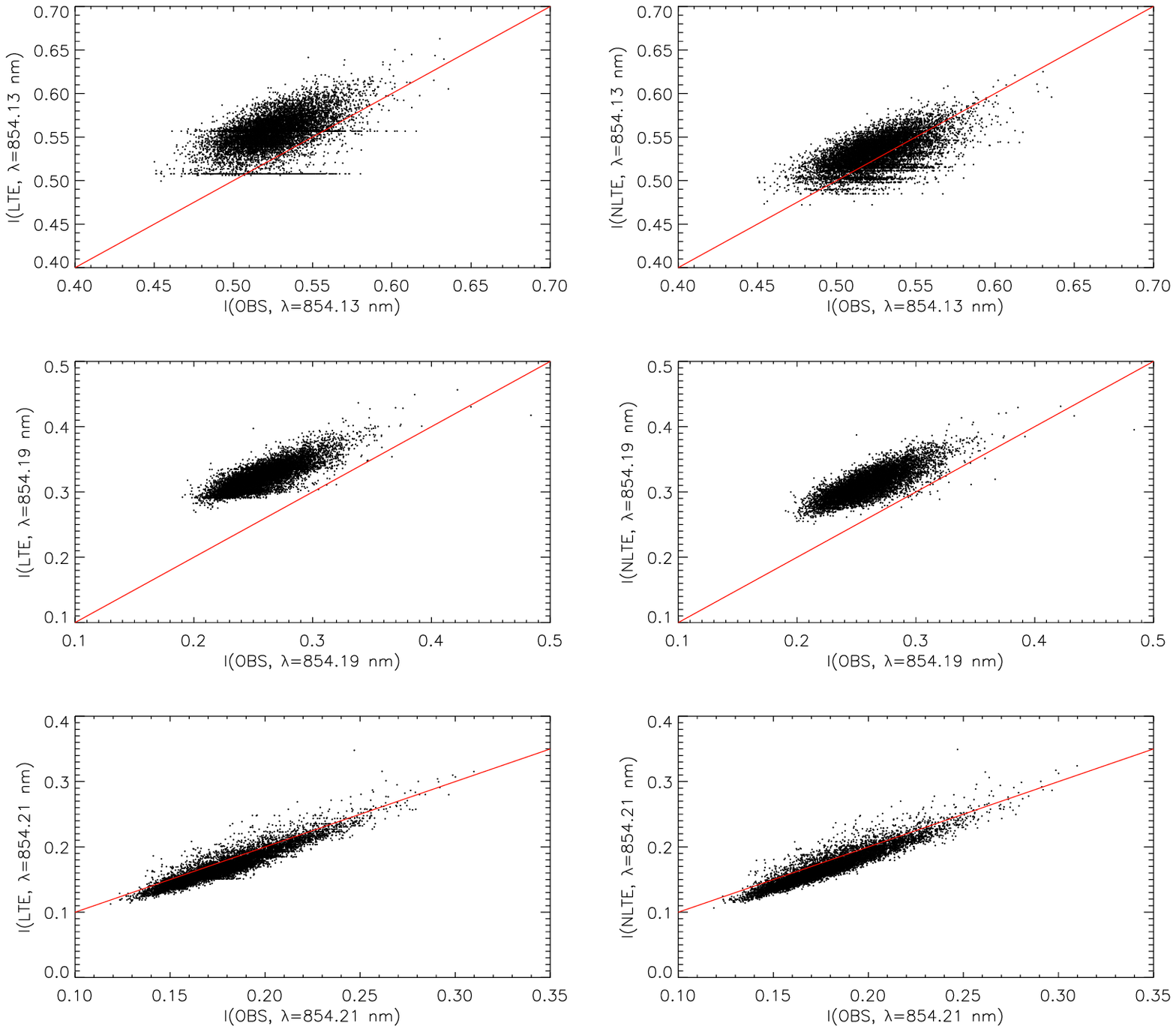}}}
\caption{Scatterplots of the intensity in the observed without Doppler shifts and best-fit spectra at three wavelengths. Left (right) column: observed vs.~best-fit LTE (NLTE) spectra. The red solid line indicates a perfect correlation.} \label{fig7}
\end{figure*}

\begin{figure}
\centerline{\resizebox{8.8cm}{!}{\includegraphics{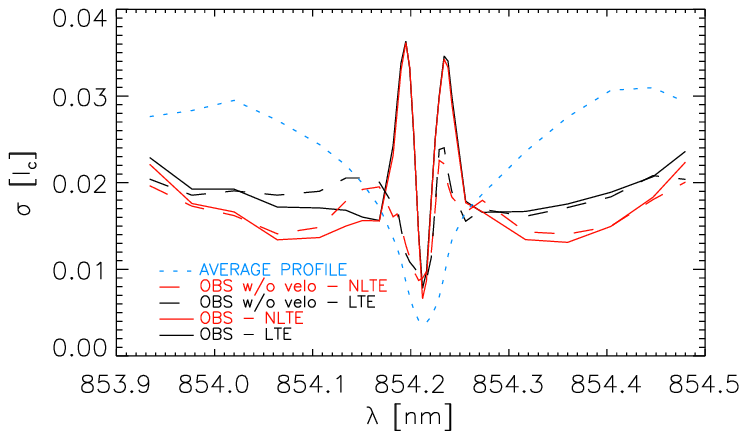}}}
\caption{Standard deviation of the residuals between observed and best-fit spectra as a function of wavelength. Black (red) solid line: standard deviation of the fit residuals for the LTE (NLTE) inversion. The dashed lines show the same for observed spectra without Doppler shifts. The blue dotted line shows the average intensity spectrum in arbitrary units for reference.} \label{fig14}
\end{figure}

Figure \ref{fig9} shows 16 randomly picked spectra including their LOS Doppler shifts and their NLTE best-fit solutions. For comparison, we also constructed a set of the observed spectra where the line-core velocities were removed. The presence of the Doppler shifts is only clearly visible for a few of the spectra, while the best-fit solution from the archive -- without any velocity -- matches both the original spectra and those without velocities satisfactorily. Neglecting the LOS Doppler shifts thus has no strong impact on the thermal stratifications retrieved because the inversion matches the line shape, i.e., the intensities in the line wing and line core.

In principle the LOS velocity can be determined and removed from the observations prior to the inversion, but for the coarse spectral sampling in the line wing in our IBIS data that approach can decrease the quality of the observed spectra. A determination of the LOS velocity from the location of the line core and a derivation using a spectral displacement of the best-fit spectra corresponding to known velocities gave about the same result for the LOS velocities. Such a step could be added to the inversion approach either prior to the inversion or as an intermediate step after one initial inversion, but it is not expected to lead to a significant improvement of the fit. We thus ran the inversion without removing the LOS velocities, but for a few of the figures below we will use the observed data set without LOS velocities because at the wavelengths with the steepest spectral intensity gradient in the profile to the blue and red of the line core the presence of the LOS velocities changes the intensity at a fixed wavelength significantly.

\begin{figure*}
\centerline{\resizebox{17cm}{!}{\includegraphics{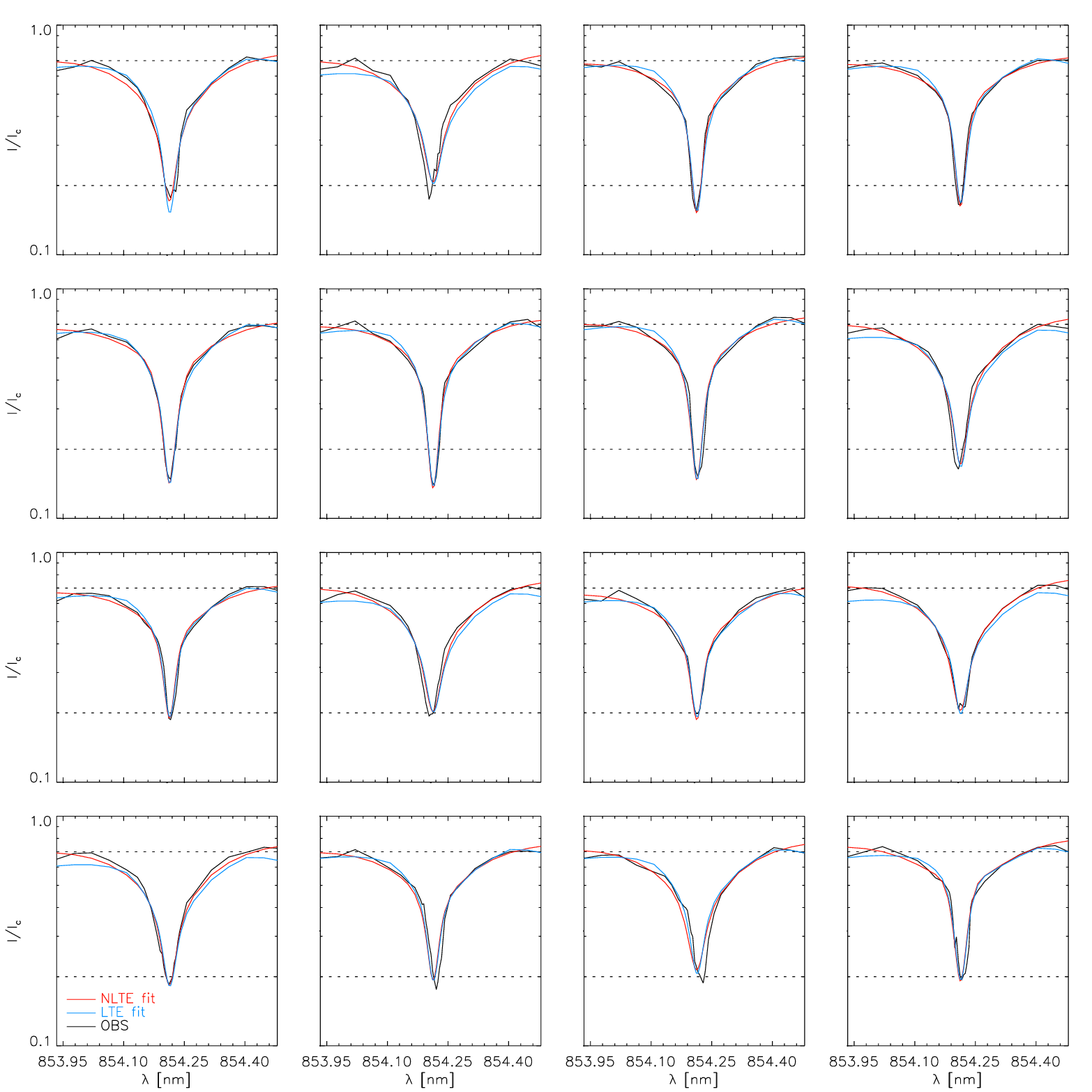}}}$ $\\$ $\\
\caption{Comparison of 16 randomly selected observed spectra including Doppler shifts (black lines) and best-fit spectra for the LTE (blue lines) and NLTE inversion (red lines).} \label{fig11}
\end{figure*}
\subsection{Fit Quality}
\subsubsection{$\chi^2$ Values}
For the investigation of the fit quality, we re-calculated the $\chi^2$ difference between the observed and best-fit profiles with equal weight for each wavelength as a better estimate of the general fit quality to the full line profile. Figure \ref{fig2} shows a two-dimensional (2D) image of the $\chi^2$ values in the LTE and NLTE inversion for one spectral scan together with the corresponding line-core image. Both the LTE and NLTE inversion have a similar range and spatial structure in $\chi^2$ with seemingly no direct dependence on the intensity pattern in the line-core image. Some locations with high values of $\chi^2$ are located on the darkest spots of the line-core image. The histograms of $\chi^2$ and the scatterplot of $\chi^2$ in the NLTE vs.~the LTE inversion in Figure \ref{fig3} confirm a similar performance of the LTE and NLTE inversion with slightly lower $\chi^2$ values for the NLTE fit. The shape of the distributions matches the expectations for a least-squares fit given the number of degrees of freedom and the number of wavelength points \citep[see][]{grant+etal2018}.

\subsubsection{Intensity at Fixed Wavelengths}
Figure \ref{fig6} shows 2D images of the observed spectra without Doppler shifts and the LTE and NLTE best-fit spectra at three wavelengths in the outer (854.13\,nm) and mid (854.19\,nm) line wing and the line core (854.21\,nm). The line core (top row of Figure \ref{fig6}) is well reproduced by both the LTE and NLTE inversion with no discernible difference between the two inversions. The same holds for the mid line wing in the middle row. The largest differences are seen for the outer line wing, where especially the LTE inversion becomes somewhat coarser in its spatial resolution. This is a consequence of lacking suited profiles for some spatial locations in the archive, so the fit is forced to use the same profile multiple times over neighboring spatial locations. The NLTE inversion maintains the spatial resolution in the line wing, but also is off from the observations to some extent, which results from both the Doppler shifts and the general problems of the observed spectra in the outer line wings caused by the coarse spectral sampling and the prefilter transmission curve. All inversion results show a slightly higher spatial contrast than the observations themselves, which is best seen in the line-core images, e.g., at the bright grain at $x,y\approx 60^{\prime\prime},25^{\prime\prime}$ in the top row of Figure \ref{fig6}.
\begin{figure*}
\centerline{\resizebox{14cm}{!}{\includegraphics{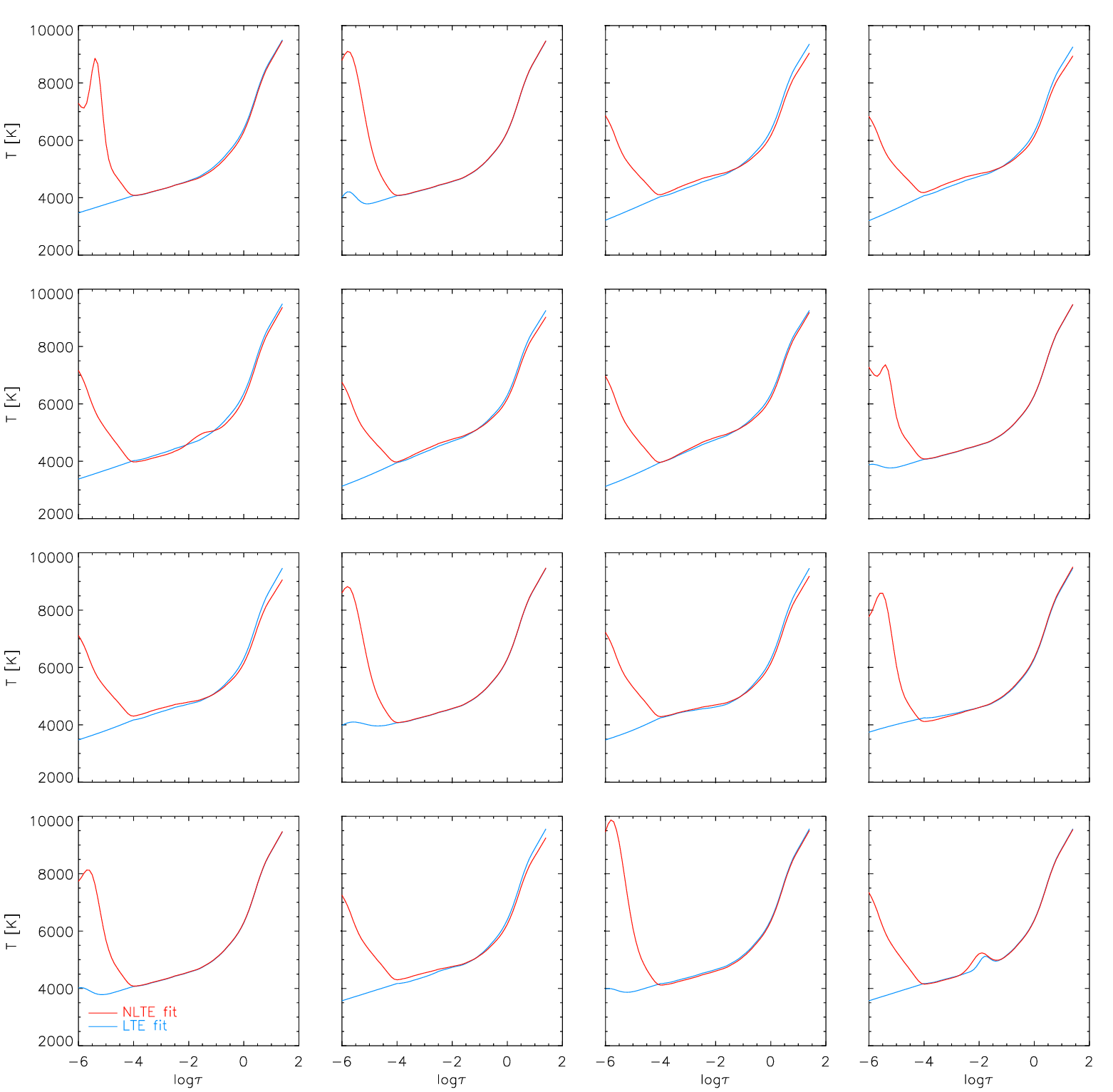}}}$ $\\$ $\\
\caption{Temperature stratifications inferred for the spectra in Figure \ref{fig11}. Blue (red) lines: LTE (NLTE) best-fit temperature.} \label{fig13}
\end{figure*}

The scatterplots of the best-fit vs.~the observed intensity at the same three wavelengths in Figure \ref{fig7} confirm the picture from the 2D images. The scatter around a one-to-one relation is smallest for the line core and increases towards the outer wing. The global offset of the points for the mid line wing in the middle row from the line of perfect correlation is due to the removal of the Doppler velocities by shifting the line-core location to a fixed reference position, which also shifts the whole spectral profile. 

The lack of suitable archive profiles for some locations is easily seen for the outer line wing in the top row. The horizontal "stripes" of identical intensity in the LTE and NLTE inversion for a range of observed intensities show that the same archive profile was used several times. For the LTE inversion it seems that this was due to a lower threshold in the intensity in the archive with no intensities below about 0.51, which implies that most likely the lowermost value of the temperature offset $\Delta T$ in its generation was not large enough.

Figure \ref{fig14} shows the standard deviation of the difference beween the observed spectra with and without Doppler velocities and the LTE and NLTE best-fit spectra. The average value of the standard deviation is about 0.02 of $I_c$ at all wavelengths. The largest values are found where the absorption profile is steepest and the difference between observed and best-fit spectra is the most sensitive to the Doppler shifts. Using the observed spectra without the line-core LOS velocities for the calculation reduces the deviation at the wavelengths in the mid line wing by a factor of about 2 to a similar value as at all other wavelengths. This implies that executing the fit with the Doppler shifts previously removed will not lead to a significantly different or better result.
\subsubsection{Individual Spectra}
Figures \ref{fig11} and \ref{fig13} show the individual observed and best-fit spectra and the resulting temperature stratifications in LTE and NLTE for another set of 16 randomly selected locations. In the spectra, the difference between the LTE and NLTE solution is negligible, implying that both archives have a comparable set of spectra suited for the inversion of the specific observation. The main difference between the LTE and NLTE solution shows up in the temperature stratifications in Figure \ref{fig13}. While the LTE and NLTE solutions are nearly identical up to log\,$\tau = -4$, only the NLTE inversion is able to reproduce the chromospheric temperature rise for higher atmospheric layers, which would lead to extreme emission profiles in the line core if treated in LTE. For cases with temperature enhancements in the low atmosphere such as in the lowermost rightmost panel, the LTE and NLTE results are very similar up to log\,$\tau = -4$.

\begin{figure}
\centerline{\resizebox{8.8cm}{!}{\includegraphics{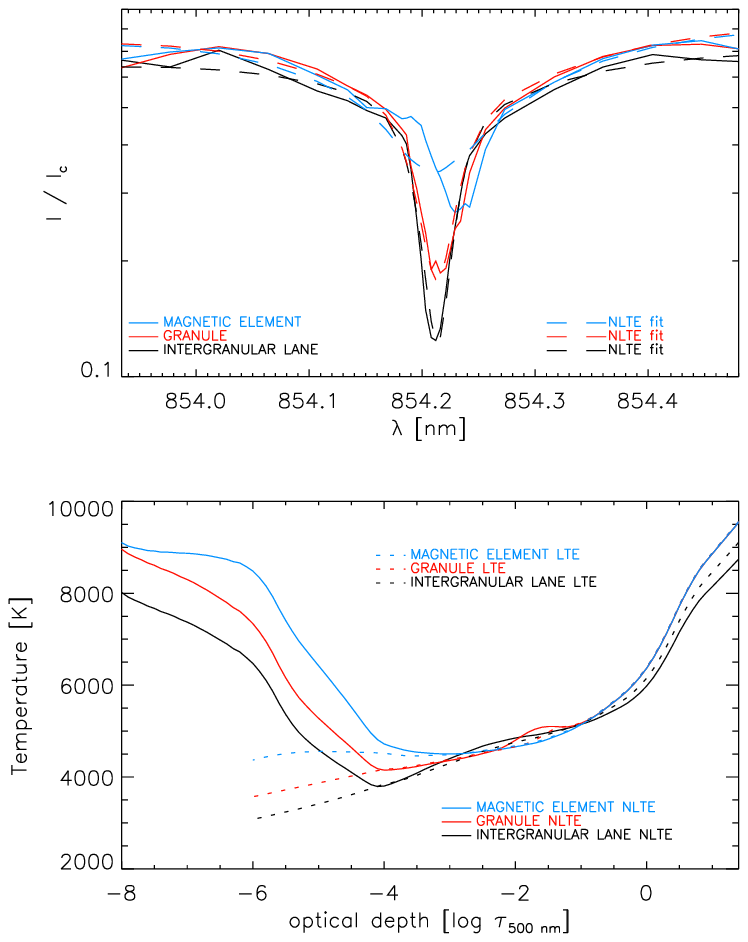}}}
\caption{Spectra (top panel) and temperature stratifications (bottom panel) for three explicitly selected spatial locations. Black lines: intergranular lane. Red lines: center of a granule. Blue lines: magnetic element. The solid and dashed lines in the top panel show the observed and NLTE best-fit spectra. The solid (dotted) lines in the bottom panel show the corresponding temperature stratifications from the NLTE (LTE) inversion.} \label{fig15a}
\end{figure}
Figure \ref{fig15a} shows the spectra and temperature stratifications for three explicitly selected locations in an intergranular lane, the center of a granule and a magnetic element. Both the spectra and the stratifications comply with the expected behavior of an increase in the intensity, especially in the very line core, and the corresponding increase in the temperature with a variation also in the slope of the temperature stratification. The LTE inversion can again only provide the corresponding relative difference in temperature without a realistic chromospheric temperature rise.
\subsection{``Small'' Archive} \label{sec_small}
The large archive size of 2,000,000 profiles is the major limitation on the speed of the NLTE inversion. However, most of those profiles are not even needed. To maximize the archive usage, we ran the inversion with the full NLTE archive over the first 60 spectral scans and retained only the about 70,000 unique archive profiles that were used at any time. This ``small'' archive was then used to invert the spectral scans Nos.~200--290 that were not used in its creation. We also inverted the spectral scans Nos.~200--210 with the full NLTE archive for comparison. 
\begin{figure*}
\centerline{\resizebox{14cm}{!}{\includegraphics{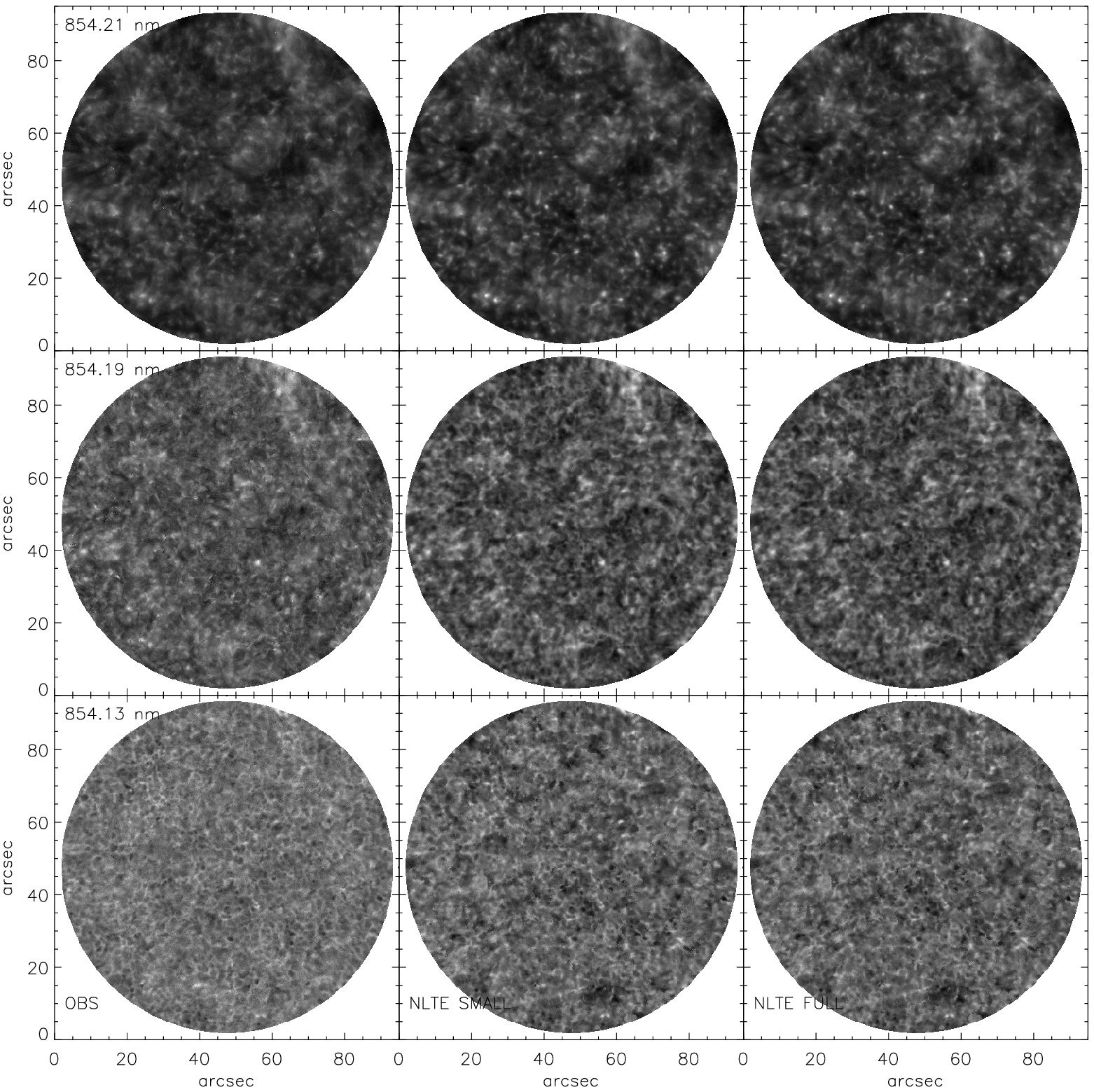}}}$ $\\$ $\\$ $\\
\caption{Comparison of 2D images of the full and small NLTE archive inversion at three wavelengths for the spectral scan No.~200. Left column: observed spectra with Doppler shifts removed. Middle and right column: best-fit spectra from the small and full-size NLTE inversion, respectively. The wavelength is indicated in the upper left corner of the panels in the first column.} \label{fig1_small}
\end{figure*}

Figure \ref{fig1_small} shows a comparison of the images at a fixed wavelength like for Figure \ref{fig6} for the spectral scan No.~200. The match of the inversion with the small archive to the observations compares well with that for the full-sized archive. The rms values of the deviations at each wavelength for the small NLTE archive stayed at a few percent similar to those in Figure \ref{fig14}. The inversion with the small archive for this specific spectral scan used 26,320 out of its 70,000 available profiles. The inversion speed improved to about 0.02\,s per profile. The full NLTE inversion used 26,523 profiles out of its 2,000,000 profiles, i.e., only about 200 more, implying that the small archive is rather complete. The comparison of the small and full NLTE inversion for the other scans Nos.~201-210 gave similar results. Table \ref{table_speed} lists the speed of the different inversion modes. Tests on a 32-core machine instead of the quad-core desktop computer gave about 10\,\% faster speeds even when running up to six inversion jobs in parallel.
\begin{table}
\caption{Summary of archive performance for spectra with 30 wavelength points.}\label{table_speed}
\centering
\begin{tabular}{cccc}
Type & Speed & Time/IBIS scan & Archive size\cr\hline\hline
LTE & 0.03\,s & 5.8 hrs& 240,000\cr
NLTE full& 0.35\,s & 68 hrs & 2,000,000\cr
NLTE small & 0.02\,s & 3.9 hrs & 70,000\cr
\end{tabular}
\end{table}
\begin{figure*}
\centerline{\resizebox{17.6cm}{!}{\includegraphics{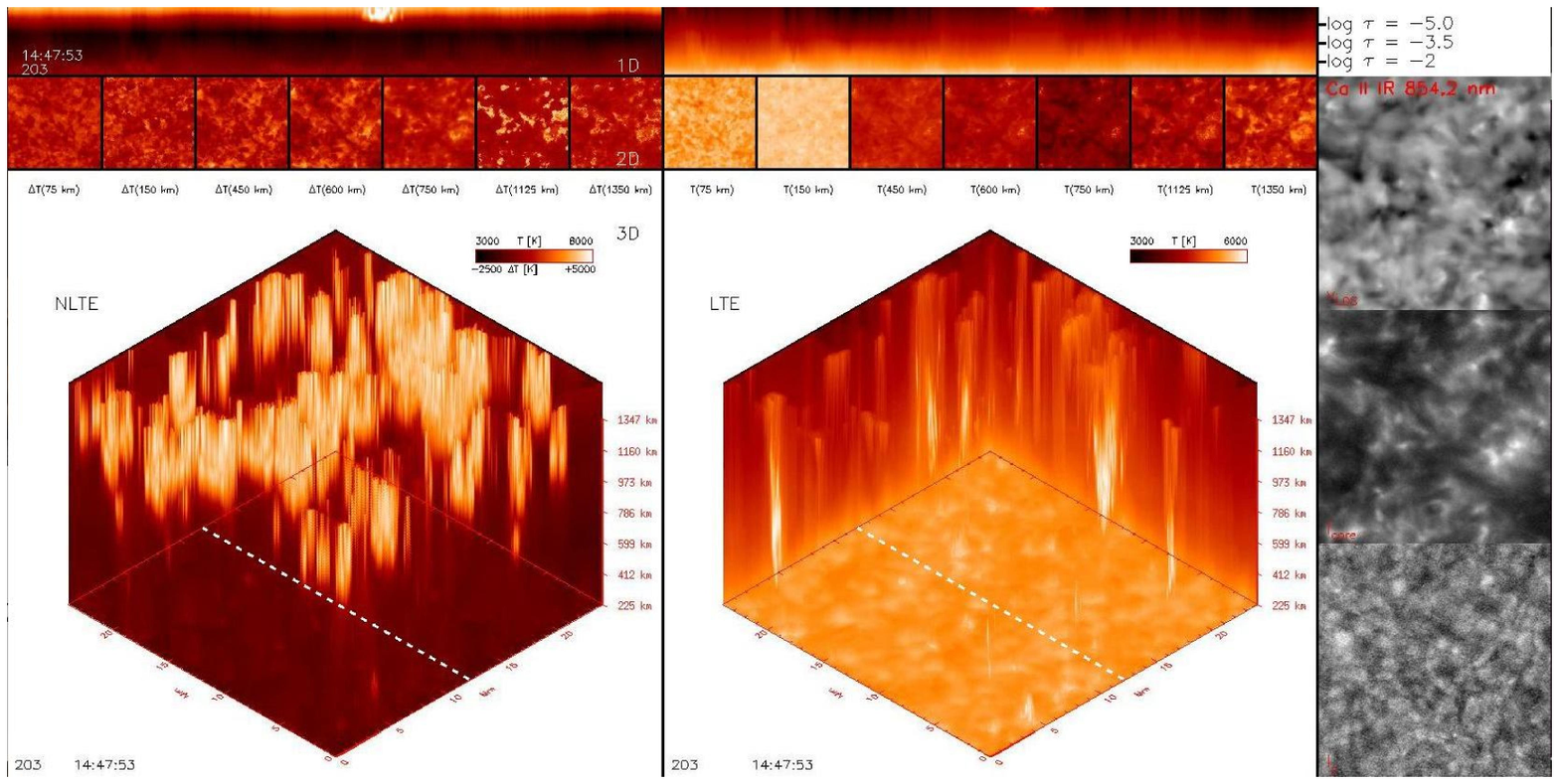}}}
\caption{Still from the animation of the inversion results for the spectral scans Nos.~200--290. Left/middle column: NLTE/LTE results for a 25\,Mm$\times$\,25\,Mm region at the center of the FOV in (top to bottom) 1D, 2D and 3D visualization. Right column, top to bottom: line-core velocity $v_{LOS}$, line-core intensity $I_{core}$ and continuum intensity $I_c$. The number of the spectral scan and its time are given in the lower left corner. The white dashed lines in the 3D plots indicate the location of the cut show in the 1D plot $T(x,\log \tau)$ at top. The 2D and 3D display of the NLTE results show the relative temperature $\Delta T = T-<T>$, while everything else shows absolute temperatures.} \label{fig2_small}
\end{figure*}
\subsection{2D and 3D Visualization of Temporal Evolution}
Figure \ref{fig2_small} shows a still from an animation of both the LTE and NLTE inversion results in a 25\,Mm\,$\times$\,25\,Mm region at the center of the IBIS FOV for the spectral scans Nos.~200--290 that cover about 15\,min of observations. The 1D display at the top shows the temperature $T(x,\log \tau)$ on a cut through the middle of the region. The 3D displays were generated in the same way as in \citet{beck+etal2014} by a ray-tracing through the temperature cube $T(x,y,z; t_i)$ obtained from the inversion for each spectral scan taken at the time $t_i$. The value to display for each ray was set to the largest temperature encountered along the ray. The conversion from $\log \tau$ to $z$ was done as in the latter publication assuming the relation between $\tau$ and $z$ given in the HSRA model as a first-order estimate. Because of the chromospheric temperature rise in the NLTE inversion results, the relative temperature $\Delta T(x,y,z)  = T(x,y,z)-<T(z)>_{|x,y}$ was used for the NLTE 3D display because otherwise always only the top layers of the atmosphere with the highest absolute temperature are seen.

When playing the animation, upwards propagating acoustic waves can be identified in the 1D display, but at the 10\,s cadence of the observations they often are captured only on a few subsequent steps. The small-scale ($< 5$\,Mm) locations of increased chromospheric temperatures at $\log \tau \approx -5$ in the 1D display are usually mapped in the same way in the LTE and NLTE inversion, similar to the locations with reduced temperature in between that extend over larger areas \citep[see, e.g.,][]{rezaei+etal2008}. 

The easiest identification of structures in the LTE 3D display is by comparing it to the line-core intensity image at the right-hand side. The structures corresponding to the strongest local brightenings in the latter can then be found in the LTE 3D display, where they reveal their shape throughout the atmosphere also below the chromospheric layers where the \ion{Ca}{ii} IR line core forms. A few of them will correspond to magnetic elements instead of acoustic shocks, but the target region at disc center was magnetically very quiet on the observing day. 

It is more difficult to then identify exactly the same structures in the NLTE 3D display because there usually are several more locations with high chromospheric temperatures that stand out. For the case of the scan No.~203 shown in Figure \ref{fig2_small}, the three most prominent bright features in the LTE 3D display at $x,y = (20,3), (20,10), (2,24)$\,Mm can be located in the NLTE 3D display as well. The majority of the other high-temperature structures in the NLTE 3D display can, on the other hand, be usually identified in the line-core intensity image as smaller and less bright intensity enhancements that presumably result from acoustic shocks \citep[``bright grains''; see, e.g.,][]{rutten+uitenbroek1991,beck+etal2008} instead of magnetic elements. Those do not show up as prominently in the LTE 3D display because their temperature in LTE is not so strongly enhanced that they dominate the ray-tracing in the absolute temperature in the same way. For the LTE 3D display, in some cases the photospheric temperatures are the largest values along a given ray. We note that this is not a flaw of the LTE or NLTE inversion results, but a generic problem in how to visualize the corresponding temperature cubes. 

In summary, the animation demonstrates that with either the LTE or NLTE inversion results chromospheric structures can be identified and traced in three dimensions throughout the atmosphere, while the application of the inversion to a time-series of observations allows one to also trace the temporal evolution in an atmosphere cube.
\section{Summary}\label{summsumm}
We generated a large archive of 2,000,000 profiles under the assumption of non-local thermodynamic equilibrium for an inversion of solar chromospheric \ion{Ca}{ii} spectra. After matching the average observed quiet Sun profile with the spectral synthesis corresponding to the unperturbed HSRA model, we applied the full archive to 60 spectral scans with 700,000 spectra each. We retained all archive spectra used in those inversions to make up a much smaller archive of 70,000 spectra that is optimized for the specific observation target of quiet Sun at disc center. We find that line-of-sight Doppler velocities have a negligible effect on the temperature stratifications retrieved. Both the full and the small archive have a comparable performance in terms of fit quality and reproduce observed spectra to within a few percent of $I_c$. The speed improved from about 0.3\,s per profile for the full NLTE archive to about 20\,ms per profile for the small NLTE archive, which allows one to invert a complete time-series with a few hundred million spectra in a few weeks with moderate computing power.
\section{Discussion}\label{disc}
We find that an archive of about 100,000 pre-calculated NLTE \ion{Ca}{ii} IR spectra is sufficient to successfully reproduce chromospheric quiet Sun observations of high spatial and spectral resolution. The optimal archive size can be reached by an application of a larger archive to only part of the data. The optimized archive allows one to reach speeds of a few ten ms per profile for inverting solar chromospheric spectra in NLTE. 

One advantage of the archive inversion approach is that there is no general restriction on the source of the temperature stratifications and the corresponding spectra. Opposite to our approach with a ``manual'' generation of the archive, one can also use the results from realistic numerical simulations as input \citep{riethmueller+etal2017} or use any other collection of plausible solar model atmospheres. One can also merge spectra from different sources as long as the general properties of the spectra such as spectral sampling and resolution are or can be made compatible.

The inversion approach is capable of dealing with multiple spectral lines at the same time \citep[e.g.,][]{riethmueller+solanki2019} because this only requires to run the initial spectral synthesis on the same temperature stratifications for all spectral lines wanted. For the case of \ion{Ca}{ii} spectra, the most promising improvements would be the inclusion of photospheric line blends \citep[e.g.,][]{felipe+etal2010,beck+etal2013} or using quasi-simultaneous observations of \ion{Ca}{ii} and photospheric lines. The main backbone of the archive are in this sense not the spectra, but the corresponding temperature stratifications.

The archive inversion approach is able to provide an automatic analysis of time-series of millions of spectra. The manual step of matching the archive to the observations has only to be executed once, while the intensity normalization of the observations to the FTS atlas in the line wing can be done in a fully automatic way. Given the speed of the inversion process, the analysis is thus not limited to individual spectral scans or snapshots, but can be run over series of spectral scans. This allows one to also study the temporal evolution in the solar chromosphere in detail \citep{beck+etal2013a}.
\subsection{Current Limitations}
The major limitation of the archive inversion approach is the match of the HSRA spectral synthesis to the average observed quiet Sun (QS) profile. This prevents a fully independent determination of absolute temperatures. All temperatures are derived in some sense as differences to the HSRA model that is forced to correspond to the average QS spectrum, and hence to first order to the average QS temperature stratification. For the scientific interpretation of the inversion results this presents, however, only a minor drawback. Firstly, the HSRA is assumed to be a valid QS temperature stratification with a small error at least for the optical depth range from $\log\tau = 0$ up to $\log \tau \approx -6$. Secondly, for most science targets no absolute reference temperature values are known from other sources, i.e., potential temperature errors of even a few 100\,K are often rather irrelevant as there is nothing to compare the inversion results to. Thirdly, relative temperature differences across the FOV are maintained in the correct way. 

The second major limitation of the approach still is its speed, even with all attempts at optimization. At 20\,ms per profile, an inversion of the full 1 hr time-series of IBIS data used in the current study takes about two months with a single job. As long as enough memory and CPU cores are available, the duration would scale linearly with the number of jobs run in parallel. That puts the approach still at the limit of practical application for data from current instruments.  

The apparent increase in spatial contrast in the inversion results presumably indicates a problem common to all inversion codes that work on a single-pixel base. The spectrum from a single spatial location is reproduced with a single thermal stratification ignoring the 3D radiative transfer in the solar atmosphere and the spatial degradation by the Earth's atmosphere and the telescope optics with lateral contributions from adjacent locations. For photospheric spectral lines in LTE with a strong coupling of radiation field and local gas temperature the single-pixel approximation in the solar atmosphere is valid, but the importance of the lateral radiation increases higher in the atmosphere \citep{leenaarts+etal2009}. This problem can be overcome by using a spatially coupled inversion approach \citep{vannoort2012} or applying a deconvolution to the observations before the inversion \citep{beck+etal2011a}, but in general it affects chromospheric spectral lines more than photospheric ones. To accurately model the 3D radiative effects in the solar atmosphere would require using an height- or wavelength-dependent spatial point spread function (PSF) in addition to the regular PSF for the spatial degradation outside the solar atmosphere.

Finally, the neglect of the LOS Doppler velocities in the inversion is a minor limitation, but they could be taken into account with little additional effort. 
\subsection{Future Work}
We currently foresee a few different avenues for future improvements of the inversion approach. 

The current full or small archive is optimized to deal with QS observations at disc center. It will only be partially suited for inversions of strongly differing solar targets such as active regions with pores and sunspots or any type of observations at large heliocentric angles. Especially for a potential application of the NLTE inversion to the full-disk spectra in \ion{Ca}{ii} IR provided by SOLIS, a set of a few archives at different heliocentric angles will be needed. Even if in the automatic generation of the archive a lot of different temperature stratifications are created, they do not necessarily resemble stratifications for inclined line of sights through the solar atmosphere. In regular operations, SOLIS \ion{Ca}{ii} IR full-disk data come in at a daily rate of one or two times about 2.7 million spectra for either Stokes $I$ and $V$ or Stokes $I, Q, U,$ and $V$ measurements. With moderate computing power, such an amount of data can be handled with the current archives overnight.

An improvement of the inversion speed might be possible by application of a Principal Component Analysis \citep[PCA; e.g.,][]{rees+etal2000,eydenberg+etal2005,casini+etal2013} to either the observed or the degraded archive spectra. Keeping only the first few principal components will reduce the dimensionality of the observed and archive spectra, which will speed up the inversion process. SOLIS spectra have about 100 wavelength points, which the PCA could reduce to less than 10 relevant parameters. The inversion process would work in a very similar way, but instead of calculating $\chi^2$ from the squared difference of observed and archive spectra it would minimize the squared difference of the PCA components. 

Finally, neural networks \citep[see, e.g.,][and references therein]{carroll+staude2001,socasnavarro2005,carroll+kopf2008} are able to reach a much faster speed than the one-by-one comparison of each observed profile with each archive spectrum. They also have the capability of additional interpolation between the thermal stratifications contained in the archive \citep{osborne+etal2019}. Preliminary tests training a neural network on the ``small'' archive (courtesy of I.~Milic) showed a tremendous increase in the inversion speed, but a more detailed discussion of this topic is outside of the scope of the current study.
\section{Conclusions}\label{concl}
An inversion approach for spectra of the chromospheric \ion{Ca}{ii} IR line at 854.2\,nm based on a spectral archive generated assuming non-local thermodynamic equilibrium can successfully reproduce observed spectra to within a few percent of $I_c$ at a speed of down to 0.02\,s per profile after optimization. Even large data sets of a few hundred million spectra can be quantitatively analyzed in a reasonable time with modest computing power. The inversion approach is able to deal with simultaneous observations of multiple spectral lines, but retrieves only the temperature stratification in hydrostatic equilibrium. 

Following the label for the corresponding LTE version, we will name the new code  ``\textit{Non-Local Thermodynamic Equilibrium CAlcium Inversion using a Spectral ARchive} (NLTE-CAISAR)''.

\begin{acknowledgements}
The Dunn Solar Telescope at Sacramento Peak/NM was operated by the National Solar Observatory (NSO). NSO is operated by the Association of Universities for Research in Astronomy (AURA), Inc.~under cooperative agreement with the National Science Foundation (NSF). We thank R.~Rezaei for the very first version of the NLTE archive. C.~Kiessner acknowledges support by the National Science Foundation's REU program through Award No.~1659878.
\end{acknowledgements}
\bibliographystyle{aa}
\bibliography{references_luis_mod}
\end{document}